\documentclass[osajnl,twocolumn,superscriptaddress]{revtex4}
\usepackage{amsmath}
\usepackage{graphicx}
\usepackage{hyperref}
\usepackage[T1]{fontenc}

\newcommand{\Avg}[1]{\langle \mathop{:} #1 \mathop{:} \rangle}
\begin{document}
\title{Joint spectrum of photon pairs measured by
coincidence Fourier spectroscopy}

\bibliographystyle{apsrev}
\author{Wojciech Wasilewski}
\address{Institute of Physics, Nicolaus Copernicus University, Grudzi\k{a}dzka 5, 87-100 Toru{\'n}, Poland}
\address{Institute of Experimental Physics, Warsaw University, Ho{\.z}a 69, 00-681 Warsaw, Poland}
\author{Piotr Wasylczyk}
\address{Institute of Experimental Physics, Warsaw University, Ho{\.z}a 69, 00-681 Warsaw, Poland}
\author{Piotr Kolenderski}
\address{Institute of Physics, Nicolaus Copernicus University, Grudzi\k{a}dzka 5, 87-100 Toru{\'n}, Poland}
\author{Konrad~Banaszek}
\address{Institute of Physics, Nicolaus Copernicus University, Grudzi\k{a}dzka 5, 87-100 Toru{\'n}, Poland}
\author{Czes{\l}aw Radzewicz}
\address{Institute of Experimental Physics, Warsaw University, Ho{\.z}a 69, 00-681 Warsaw, Poland}

\begin{abstract}
We propose and demonstrate a method for measuring the joint spectrum of photon pairs via Fourier spectroscopy. The
biphoton spectral intensity is computed from a two-dimensional interferogram of coincidence counts. The method has been
implemented for a type-I downconversion source using a pair of common-path Mach-Zender interferometers based on Soleil
compensators. The experimental results agree well with calculated frequency correlations that take into account the
effects of coupling into single-mode fibers. The Fourier method is advantageous over scanning spectrometry when
detectors exhibit high dark count rates leading to dominant additive noise.
\end{abstract}

\ocis{270.5290, 190.4410, 300.6300}
\maketitle

Correlated pairs of photons are a popular choice in efforts to implement emerging quantum-enhanced
technologies. Proof-of-principle experiments have demonstrated ideas such as quantum
cryptography,\cite{GisinRMP02} quantum clock synchronization,\cite{Giovannetti01,ValenciaAPL04} quantum optical
coherence tomography,\cite{NasrPRL03} and one-way quantum computing.\cite{Walther05} In parallel with the expanding
range of potential applications, the need to develop appropriate tools to engineer and to characterize sources of
photon pairs is becoming apparent. Among various degrees of freedom describing optical radiation, the spectral one is
essential to a number of techniques.\cite{Giovannetti01,ValenciaAPL04,NasrPRL03}
 Also in other protocols, based on degrees of freedom such as polarization,\cite{Walther05} the spectral
characteristics needs to be carefully managed in order to ensure the required multiphoton interference
effects. This demand has brought a number of methods to control the spectral properties of photon pairs by engineering
nonlinear media and the pumping and the collection
arrangements.\cite{KurtsieferPRA01,URenQIC03,KonigAPL04,TorresOL05,LeePRA05} A development that is needed to match
these advances is the ability to diagnose accurately two-photon sources and to measure reliably their characteristics.
An important work in this context is the recent application of scanning spectrometers to obtain joint spectra of photon
pairs.\cite{KimOL05}

In this paper we demonstrate experimentally two-photon Fourier spectroscopy as a method to measure the joint
spectrum of photon pairs. The setup is based on two independently controlled Fourier spectrometers in the
common-path configuration. Such an arrangement guarantees long temporal stability necessary to characterize weak sources of
radiation operating at single-photon levels. We show that the two-dimensional map of coincidence counts recorded as a
function of delays in two interferometers can be used to reconstruct the joint spectrum of photon pairs.
We present a measurement for a type-I spontaneous down-conversion process in a bulk beta-barium borate (BBO) crystal, and
compare the results of the reconstruction with theoretical predictions.

\begin{figure}[b]
    \begin{center}
        \includegraphics[scale=0.28]{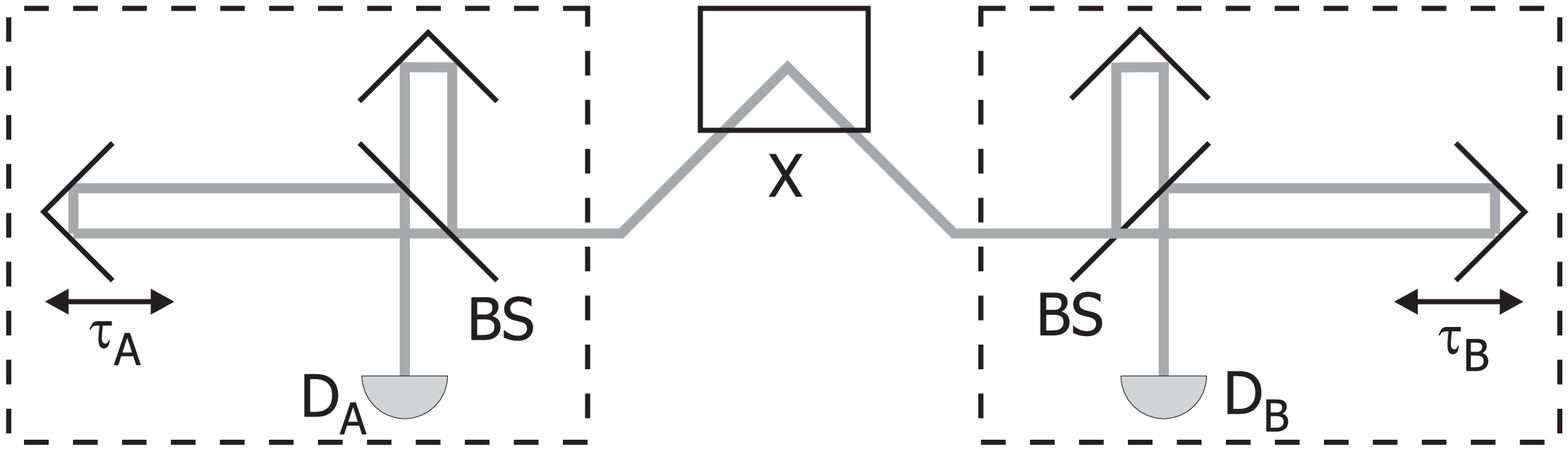}
        \caption{A simplified scheme of the double Fourier spectrometer applied to a source X of photon pairs.
        The photons enter interferometers with independently adjusted optical delays $\tau_A$ and $\tau_B$
         and are counted by detectors $\text{D}_A$ and $\text{D}_B$.}
        \label{fig:interf}
    \end{center}
\end{figure}

An idealized scheme of the experimental method is presented in Fig.~\ref{fig:interf}. The source X produces
nondegenerate pairs of photons in distinct spatial modes represented by diverging lines. Each photon is sent into a
separate interferometer, where it is divided on a beam splitter BS into two wavepackets subjected to a delay difference
$\tau_A$ and $\tau_B$ for the photon $A$ and $B$. The two wavepackets interfere on a beamsplitter BS and light emerging
from the output of the interferometer is detected. A coincidence event is recorded if both photons reach their
respective detectors. The quantity of interest is the coincidence probability  $p_{AB}(\tau_A,\tau_B)$ measured as a
function of the delays $\tau_A$ and $\tau_B$.

\begin{figure}
    \begin{center}
        \raisebox{3cm}{(a)}\hskip0mm\includegraphics[scale=0.67]{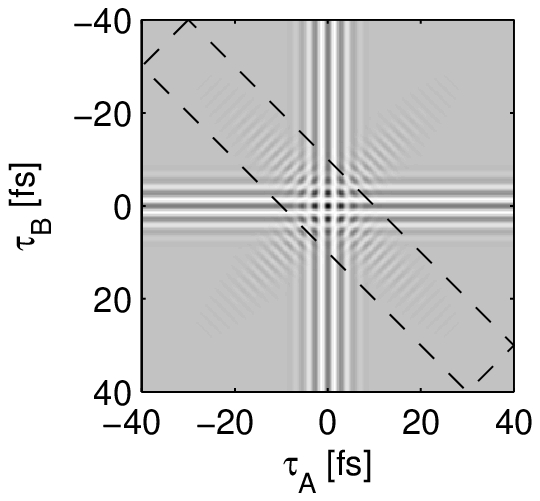}
        \raisebox{3cm}{(b)}\hskip0mm\includegraphics[scale=0.6]{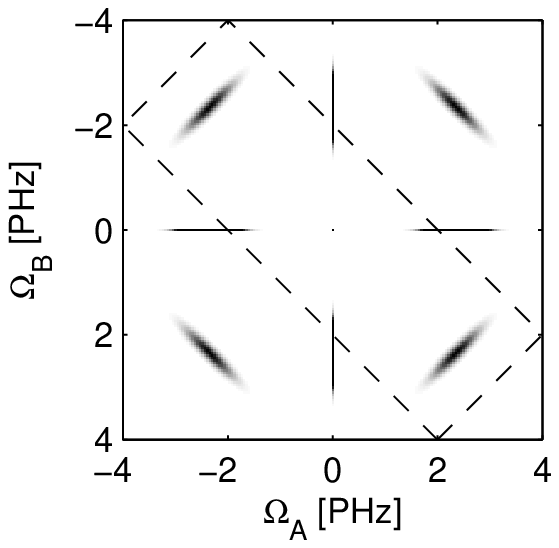}
        \caption{(a) A typical coincidence interferogram as a function of optical path differences $\tau_A$ and $\tau_B$
        and (b) its Fourier transform. The dashed rectangle in (a) defines the scan range, which with suitable sampling
        density yields the region of interest in the frequency domain, outlined in (b).}
        \label{fig:coinc}
    \end{center}
\end{figure}

The measurement is carried out in the regime when the response time of the detector is much longer than the inverse of
smallest bandwidth characterizing the spectrum of the source. Then the only relevant characteristics of the source
is the joint spectral intensity given by $\langle \mathop{:} \hat{I}_A(\omega_A) \hat{I}_B(\omega_B) \mathop{:}
\rangle$, where $\hat{I}_i(\omega_i)$ ($i=A,B$) are the spectral intensity operators for the beams $A$ and $B$, and
$\langle:\ldots:\rangle$ denotes the quantum mechanical expectation value of normally ordered operators. For
a photon with a well defined frequency $\omega$, the probability of reaching the detector is given by the standard
expression $(1+\cos \omega\tau_i)/2$. Consequently, the probability $p_{AB}(\tau_A,\tau_B)$ of a coincidence event for
a source with an arbitrary spectrum takes the form:
\begin{eqnarray}\label{eq:pAB}
p_{AB}(\tau_A,\tau_B)& \propto & \frac{1}{4} \int d\omega_A \int d\omega_B \,
\langle \mathop{:} \hat{I}_A(\omega_A) \hat{I}_B(\omega_B) \mathop{:} \rangle \nonumber\\
&&\quad\times(1+\cos \omega_A\tau_A)(1+\cos \omega_B\tau_B)
\end{eqnarray}
An example of the coincidence interferogram is shown in Fig.~\ref{fig:coinc}(a).
The two-dimensional Fourier transform of the coincidence probability
$p_{AB}(\tau_A,\tau_B)$ comprises the following terms:
\begin{eqnarray}\label{eq:FFTpAB}
\lefteqn{\int d\tau_A d\tau_B\, p_{AB}(\tau_A,\tau_B) \exp(i\Omega_A \tau_A +i \Omega_B \tau_B)
 } \nonumber \\
&\propto&\delta(\Omega_A) \delta(\Omega_B) \Avg{\hat{N}_A \hat{N}_B} \nonumber \\
&&+\frac{1}{2} \delta (\Omega_A) \Avg{\hat{N}_A \hat{I}_B(|\Omega_B|)}
+\frac{1}{2} \delta (\Omega_B) \Avg{ \hat{I}_A(|\Omega_A|) \hat{N}_B} \nonumber \\
&&+\frac{1}{4} \Avg{ \hat{I}_A(|\Omega_A|) \hat{I}_B(|\Omega_B|)}
\label{Eq:FT}
\end{eqnarray}
where $\hat{N}_i = \int d\omega \, \hat{I}_i(\omega)$ is the operator of the total photon flux in the $i$th beam,
$i=A,B$. The first term, localized at $\Omega_A=\Omega_B=0$, is proportional to the total number of photon pairs. The
two middle terms lie on the axes $\Omega_A=0$ or $\Omega_B=0$, and have the shape of the single-photon spectra
conditioned upon the detection of the conjugate photon. Finally, the last term contains the sought joint two-photon
spectrum. For optical fields, these terms occupy distinct regions in the $\Omega_A, \Omega_B$ plane and can be easily
distinguished, as shown in Fig.~\ref{fig:coinc}(b). It is helpful to trace the origin of the four terms on the right
hand side of Eq.~(\ref{Eq:FT}) to the coincidence interferogram.  The vertical and the
horizontal fringes generate the terms $\delta (\Omega_A) \Avg{\hat{N}_A \hat{I}_B(|\Omega_B|)}$ and $\delta (\Omega_B)
\Avg{ \hat{I}_A(|\Omega_A|) \hat{N}_B}$, whereas it is the diagonal fringe pattern that contains information about the
joint spectrum $\Avg{\hat{I}_A(|\Omega_A|) \hat{I}_B(|\Omega_B|)}$. This defines the region of the coincidence
interferogram that needs to be scanned in order to compute the joint spectrum. It has the shape of a tilted rectangle
outlined in Fig.~\ref{fig:coinc}(a). Let us note that the grid spacing in a given direction
can be adjusted to the characteristic scale of interferogram structures. Specifically, the grid can be sparse
in the direction parallel to the fringes, while in the perpendicular direction it needs to be fine enough to resolve
the oscillations. Then the Fourier transform of the experimental data covers the region marked with a dashed rectangle in
Fig.~\ref{fig:coinc}(b), containing the joint spectrum.

\begin{figure}
\begin{center}
    \begin{tabular}{cc}
    \raisebox{2cm}{(a)} & \includegraphics[width=0.35\textwidth]{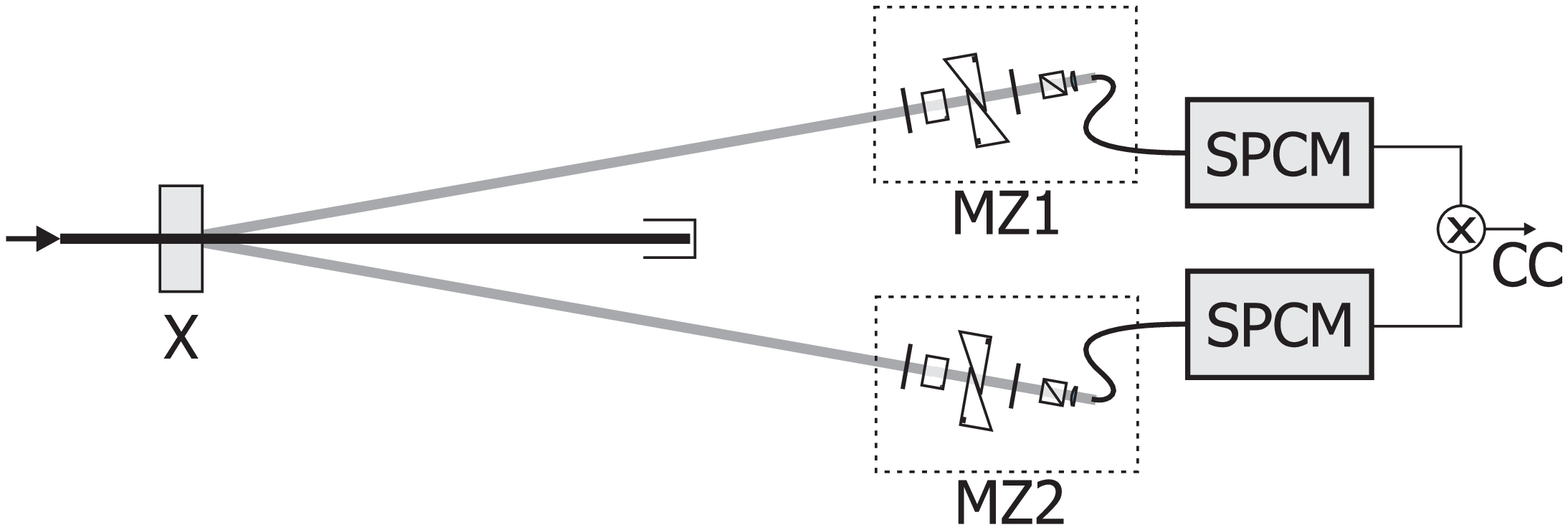}\\
    \raisebox{2cm}{(b)} & \includegraphics[width=0.3\textwidth]{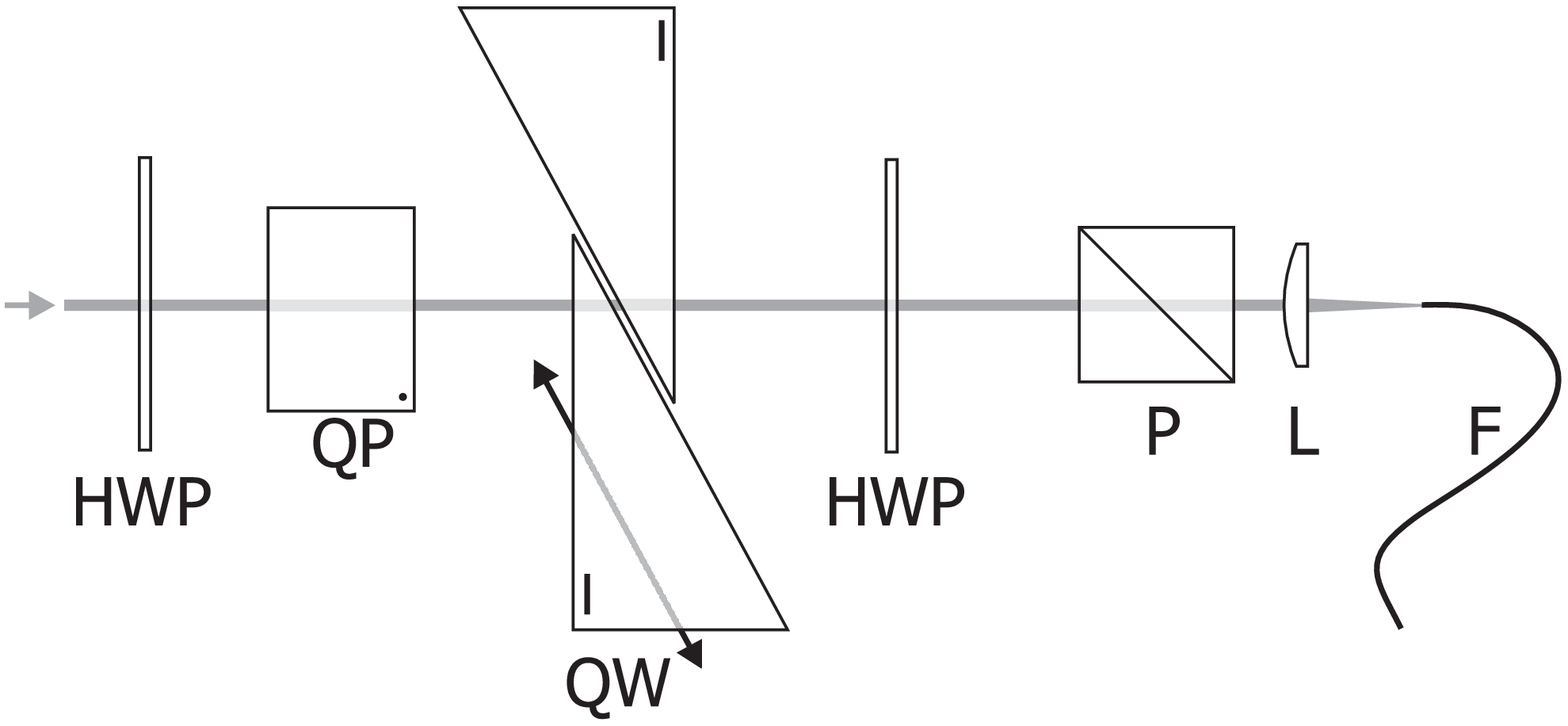}
    \end{tabular}
        \caption{(a) Experimental setup. X, BBO crystal; MZ1, MZ2, Mach-Zender interferometers;
        SPCM, single photon counting module.
        (b) Common-path Mach-Zender interferometer. HWP, half-wave plate; QP, quartz plate; QW,
        quartz wedges; P, polarizer; L, aspheric lens; F, single-mode fiber.}
    \label{fig:setup}
\end{center}\end{figure}

Our experimental setup is depicted in the Fig.~\ref{fig:setup}(a). The photon pairs were generated in a 1~mm thick
nonlinear BBO crystal in a type-I process. The crystal was pumped by 100~fs long pulses centered at 390~nm, 20~mW
average power and a repetition of 80~MHz. The ultraviolet beam was focused on the crystal to a spot measured to be
155~$\mu$m in diameter. The crystal was cut at $29.7^\circ$ to the optic axis, and oriented perpendicular to the pump
beam. Two Mach-Zender interferometers MZ1 and MZ2 collected down-converted light at angles $1.28^\circ$ and
$1.05^\circ$. The photons transmitted through the interferometers were coupled into single-mode fibers defining the
spatial modes in which the down-conversion is collected.\cite{Dragan2004} Finally the photons were detected using
single photon counting modules SPCM connected to fast coincidence electronics and a PC controlled counter board.

In order to ensure the stability of the interferometric setup over the entire two-dimensional scan we used a pair of
common path Mach-Zenders, in which the two arms were implemented as orthogonal polarization components while the
optical path difference was modulated by a Soleil compensator, as depicted in Fig.~\ref{fig:setup}(b). The generated
photons entered from the left, with their polarization set to $45^\circ$ by the half-wave plate HWP. Then the photons
went through a block of crystalline quartz QP with a vertical optic axis, and a pair of wedges QW with horizontal optic
axes. With this setup, the sign and the value of the delay between the horizontal and vertical polarization components
could be accurately controlled by sliding one of the wedges, mounted on the stepper motor driven translation stage,
into the beam. Finally the two polarizations were brought to interference using a second half-wave plate HWP and a
polarizing beam-splitter P, and the transmitted photons were coupled into a single-mode fiber F using aspheric lens L
and detected. We verified that for the spectral range of interest birefringence dispersion could be neglected and
consequently the delays $\tau_i$ were linearly dependent on the displacements of the quartz wedges. The quartz blocks
used in the setup allowed to vary the delays from $-250$~fs to 250~fs.

\begin{figure}
    \begin{center}
    \begin{tabular}{rcrc}
        \includegraphics[scale=0.72]{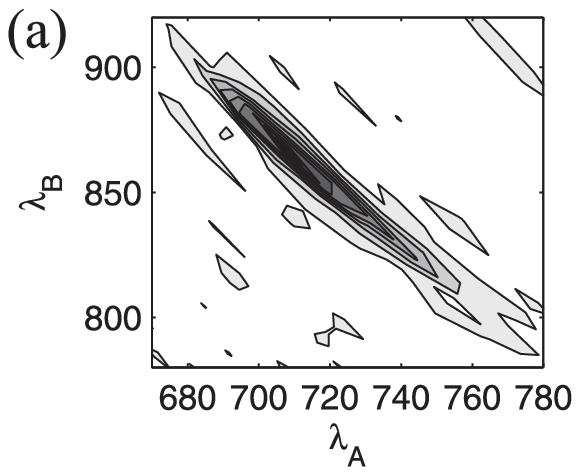} &
        \includegraphics[scale=0.72]{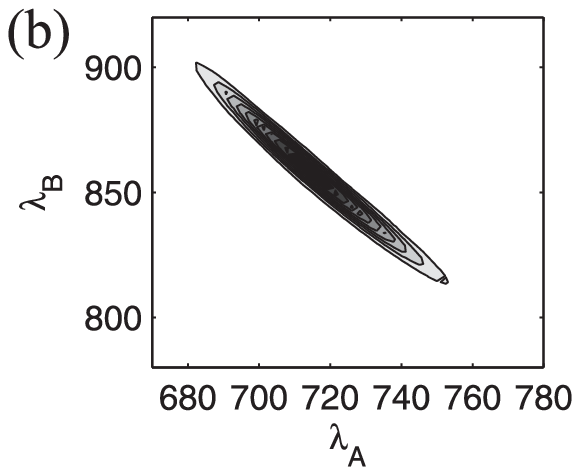} \\
        \includegraphics[scale=0.7]{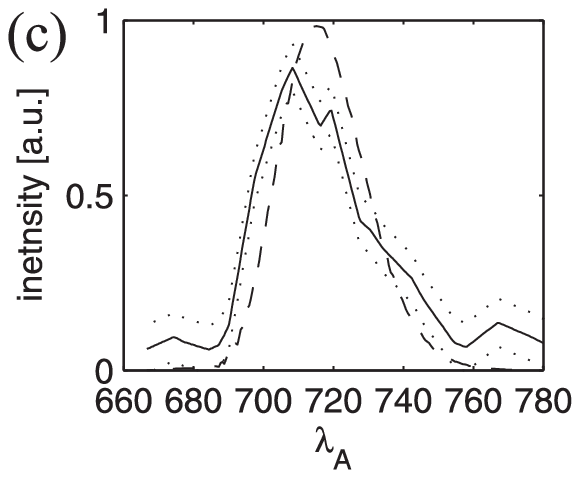} &
        \includegraphics[scale=0.7]{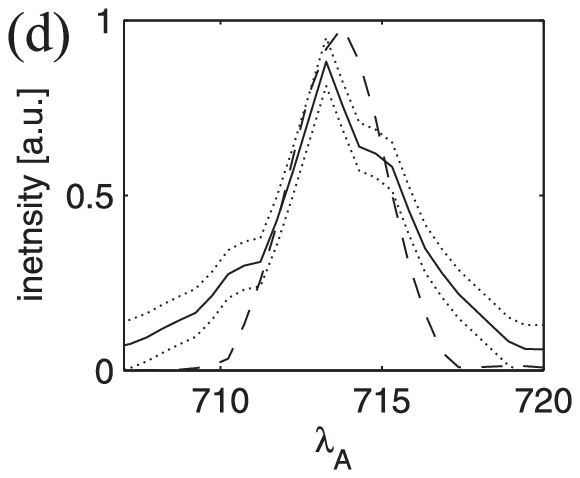}
    \end{tabular}
        \caption{(a) Measured joint spectrum of photon pairs $\Avg{ \hat{I}_A(\lambda_A)
        \hat{I}_B(\lambda_B)}$ as a function of the wavelengths $\lambda_A$ and $\lambda_B$;
        (b) theoretical predictions of the joint spectral intensity; the
        cross-sections of the theoretical (dashed line) and the
        experimental (solid with dots marking errors) distributions along
         the lines (c) $\lambda_A^{-1}+\lambda_B^{-1}=\text{const}$ and
        (d) $\lambda_A^{-1}-\lambda_B^{-1}=\text{const}$ passing through the maximum.} \label{fig:result}
    \end{center}
\end{figure}

The complete measurement consisted in a scan of a rectangular grid depicted in Fig.~\ref{fig:interf}(a) spanned by
800$\times$40 points, where the first number refers to the direction perpendicular to the fringes. The corresponding
mesh was 0.57~fs $\times$ 2~fs, and coincidences were counted for 3~s at each point. The reconstructed joint spectrum
of photon pairs, normalized to unity, is depicted in Fig.~\ref{fig:result}(a). We compare it with theoretical
calculations plotted in Fig.~\ref{fig:result}(b). The theoretical model used in these calculations assumed the exact
phase matching function of the nonlinear crystal. The transverse components of the wave vectors for the pump and
down-converted beams were treated in the paraxial approximation. The joint spectrum was calculated for coherent
superpositions of plane-wave components of the down-conversion light that add up to localized spatial modes defined by
the collecting optics and single-mode fibers. In order to facilitate a more quantitative comparison,
Figs.~\ref{fig:result}(c) and (d) show the cross sections of the joint spectra along directions of maximum and minimum
width in the frequency domain. In these plots, the experimental data have been interpolated between the points of the
Fourier-domain grid, and presented together with statistical errors calculated assuming Poissonian noise affecting
coincidence counts.

In summary, we proposed Fourier spectroscopy for measuring the joint spectrum of photons pairs, and demonstrated its
application to down-converted light generated in a type-I BBO crystal. The result of the reconstruction agrees well
with a careful theoretical calculation of the joint spectrum. We were able to reduce substantially the overall duration
of the measurement by selecting the region of the interferogram which contains information about the relevant
characteristics of the spectrum. Finally, let us note that compared to scanning spectrometers, Fourier
spectroscopy exhibits higher signal-to-noise ratio when detection noise is dominated by an additive contribution. This
effect, known as the multiplex advantage,\cite{TaiAO76} is important in the case of high dark count rates, which is
typical for single-photon measurements performed at telecom wavelengths.\cite{GisinRMP02}

This research was in part supported by KBN grant number 2P03B 029 26 and it has been carried out in the National
Laboratory for Atomic, Molecular, and Optical Physics in Toru\'{n}, Poland. P.W. gratefully acknowledges the support of
the Foundation for Polish Science (FNP) during this work.

\newcommand{\urlprefix}{ }
\renewcommand{\url}[1]{}

\end{document}